\documentclass[fleqn,usenatbib]{mnras}

\usepackage{newtxtext,newtxmath}
\usepackage[T1]{fontenc}

\DeclareRobustCommand{\VAN}[3]{#2}
\let\VANthebibliography\thebibliography
\def\thebibliography{\DeclareRobustCommand{\VAN}[3]{##3}\VANthebibliography}

\usepackage{graphicx}	% Including figure files
\usepackage{amsmath}	% Advanced maths commands
\title[Transient QPOs in MAXI J1535$-$571]{Wavelet analysis of the transient QPOs in MAXI J1535$-$571 with \textit{Insight-HXMT}}

\author[X. Chen et al.]{
X. Chen,$^{1,2}$
W. Wang,$^{1,2}$\thanks{E-mail: wangwei2017@whu.edu.cn}
P F. Tian,$^{1,2}$
P. Zhang,$^{1,2}$
Q. Liu,$^{1,2}$
H. J. Wu,$^{1,2}$
N. Sai,$^{1,2}$
Y. Huang,$^3$
L. M. Song,$^3$
\newauthor
J. L. Qu,$^3$
L. Tao,$^3$
S. Zhang,$^3$
F. J. Lu,$^3$
S. N. Zhang$^3$
\\
$^{1}$Department of Astronomy, School of Physics and Technology, Wuhan University, Wuhan 430072, China\\
$^{2}$WHU-NAOC Joint Center for Astronomy, Wuhan University, Wuhan 430072, China\\
$^{3}$Key Laboratory of Particle Astrophysics, Institute of High Energy Physics, Chinese Academy of Sciences, Beijing 100049, China
}

\date{Accepted XXX. Received YYY; in original form ZZZ}

\pubyear{2022}

\begin{document}
\label{firstpage}
\pagerange{\pageref{firstpage}--\pageref{lastpage}}
\maketitle

\begin{abstract}
Using wavelet analysis and power density spectrum, we investigate two transient quasi-periodic oscillations (QPOs) observed in MAXI J1535$-$571 observed with \textit{Insight}-HXMT. The transient QPOs have a centroid frequency of $\sim 10$ Hz with a FWHM $\sim 0.6$ Hz and an rms amplitude $\sim 14\%$. Energy spectra of QPO and non-QPO regimes are also separated and analyzed, and the spectra become softer with higher $E_{cut}$ in the non-QPO regime compared to the QPO regime. Our results suggest that the transient QPOs detected in MJD 58016 and 58017 are still the type-C QPO, and the source remains in its HIMS. The duration of all type-C QPO signals based on wavelet is positively correlated with the mean count rate above $\sim 10$ keV, implying appearance of QPOs in different time scales should be coupled with the corona. The transient QPO properties could be related to the jet or flares, perhaps the partial ejection of the corona is responsible for the disappearance of the type-C QPO.

\end{abstract}


%%%%%%%%%%%%%%%%%%%%%%%%%%%%%%%%%%%%%%%%%%%%%%%%%%

%%%%%%%%%%%%%%%%% BODY OF PAPER %%%%%%%%%%%%%%%%%%

\section{Introduction}

Black hole (BH) low-mass X-ray binaries (LMXBs) consist of a black hole accreting material from a low-mass comparison star via Roche-lobe overflow. The temperature of accreting material near the BH is high enough to radiate X-rays which provide a valid way to study the accretion disc and flow. LMXBs can be classified into transient and persistent sources. Most of the LMXBs in the BH systems are transients \citep{Fender2012}, they only go to outbursts occasionally, and spend most of their life in quiescence. During an outburst, BH LMXBs evolve through different states, normally including the hard state, the soft state, and two intermediate states. A 'q'-shape evolution routine can be noticed in the hardness-intensity diagram.

Although quasi-periodic oscillations (QPOs) are very common in the X-ray light curves, their physical origin is still ambiguous. Generally, QPOs can be classified into Low frequency (LF) QPOs and high frequency (HF) QPOs, based on the centroid frequency. LF QPOs are very common in the BH systems with its centroid frequency $\lesssim$ 30 Hz, while the HF QPOs are very rare in the BH systems with centroid frequency greater than $\sim$ 60 Hz \citep{Belloni2010}. There are three types of LF QPOs, type-A, -B and -C \citep{Casella2005}. Type-C QPOs are the most common type so far, and can be detected in almost all accretion states \citep{Munoz2014}. Its centroid frequency is from a few mHz to $\sim$10 Hz, with a high-amplitude and narrow peak and a quality factor larger than $\sim 8$. Type-B QPOs only appear in the soft intermediate state (SIMS), with a lower amplitude of $\lesssim$5\% and narrow peak of the centroid frequency around 5-6 Hz. Type-A QPOs are very rare, and they generally appear in the soft state as a very weak and broad peak and a quality factor $\lesssim 3$.

The most commonly used technique for QPO study so far is to fit the power density spectrum (PDS) with a multi-Lorentzian function \citep{Belloni2002}. In order to increase the credibility, the light curves are generally separated by time segments of the same length. Performing PDS on the separated light curves and then averaging provide the final result. Apparently, this method hide the relationship of QPO evolution over time, thus the dynamical PDS are often used to study the sudden appearance and disappearance of QPOs \citep{Huang2018,Xu2019}. However, dynamical PDS is inefficient \citep{Torrence1998} and affected by the selected time window \citep{Kaiser2011}. Recently, \cite{Chen2022} used the wavelet analysis to study the QPO evolution. With the higher time-frequency domain accuracy of the wavelet results, they found the QPO appearance/disappearance in the order of seconds. This fast evolution and variation cannot be resolved by the dynamical PDS, since the time window used in the later is normally more than a few tens of seconds \citep[e.g.][]{Zhang2021}.

The X-ray spectrum of a BH LMXB normally can be fitted with two or three components, i.e. a disc component at lower energy, a corona component at higher energy, and a possible reflection component. The rms spectrum and the phase-lag spectrum of the QPO can provide extra information to study the relation of the above components and QPO generation. LF QPOs are generally considered to be mainly related to the corona, because their rms variability increase with energy below 30 keV \citep[e.g.][]{Rodriguez2002,Yadav2016,Zhang2017,Zhang2020}, and the QPO rms spectrum at high energy \citep[$\sim$ 200 keV;][]{Ma2021} cannot be explained by the reflection part. Hard lags are believed to be related to the inverse Compton scattering inside the corona \citep{Miyamoto1988}, while soft lags may be caused by the re-emitted upscattered photons from the disc \citep{Lee1998}. The inclination dependence of QPOs \citep{Schnittman2006,Motta2015,Heil2015} suggests a geometric origin, thus the Lense-Thirring precession model \citep{Ingram2009} is commonly accepted \citep[e.g.][]{Ingram2016,Nathan2022} to explain the rms amplitude and lags of the QPOs. Recently, a time-dependent Comptonization model is proposed by \cite{Karpouzas2020} to study the energy-dependent rms and phase lag spectrum in the scenario of kilohertz QPOs in the neutron star system, and is then modified by \cite{Bellavita2022} to be suitable for a BH system. This model has been applied to study the type-C QPO in GRS 1915+105 \citep{Karpouzas2021,Garcia2022,Mendez2022} and in MAXI J1535$-$571 \citep{Zhang2022}, and the type-B QPO in MAXI J1348$-$630 \citep{Garcia2021}.

From the hard intermediate state (HIMS) to SIMS transition, the fractional rms variability integrated over a broad range of frequencies normally decreases rapidly, along with the transition of QPO type from type-C to -B \citep{Wijnands1999,Homan2001,Remillard2002,Casella2005}. This X-ray transition is accompanied by a change in the radio jet, i.e. the steady, compact jet quenching \citep{Corbel2013,Russell2011,Russell2014} and the transient jet launching compared with bright radio flares \citep{Terarenko2017b}. Up to date, only a few transient jets have been directly resolved in the BH LMXB systems \citep{Mirabel1994,Hjellming1995,Tingay1995,Fender1999,Mioduszewski2001,Yang2010,Miller-Jones2019}, thus the physical origin of the jet is still unclear, and no clear evidence regarding the association of the discrete, transient jet and the type-B QPO has been detected \citep{Fender2009,Miller-Jones2012}. The jet ejection could be related to some other effect, such as the rapid decrease in X-ray rms, the sudden increase of X-ray count rate, or the change of the centroid frequency of the type-C QPOs \citep{Russell2019}.

MAXI J1535$-$571 is an X-ray transient discovered by \textit{MAXI}/GSC \citep{Negoro2017a} and \textit{Swift}/BAT \citep{Kennea2017b} independently during its 2017 outburst. Follow-up observations suggested that the source is a BH X-ray binary \citep{Negoro2017b,Russell2017} with a high spin and high inclination \citep{Miller2018,Xu2018}. The source went into the HIMS about one week after its discovery \citep{Kennea2017a,Nakahira2017,Palmer2017,Shidatsu2017b,Tetarenko2017a}, and transitioned into the soft state after one or two months \citep{Shidatsu2017a}. During the intermediate state, LF QPOs of both type-C and -B were detected \citep{Gendreau2017,Mereminskiy2017,Huang2018,Stevens2018,Bhargava2019,Vincentelli2021}.

In this paper, wavelet analysis and PDS methods are used to study two transient QPOs in the source MAXI J1535$-$571 during its 2017 outburst with \textit{Insight}-HXMT data. We describe the observations and data reductions in Section 2, and explain how the QPO phase-lags are calculated with the cross spectra. In section 3, we show the timing results obtained with different techniques, along with the energy and time dependence of the QPO properties. The energy spectrum fitting results are presented in Section 4, comparing the differences between the spectra of intervals with and without QPO (hereafter QPO spectra and non-QPO spectra, respectively). In Section 5, we discuss our results and compare the X-ray data with the radio flares/jet reported in the former research, and a brief conclusion is summarized in Section 6.

%We also show the evolution of the S-factor of 11 observations in which the QPO phenomenon was detected in the \textit{Insight}-HXMT observations, and analyze the appearance duration and intensity of the QPO signals in these observations.

\section{Observations \& Data Analysis}

The hard X-ray Modulation Telescope (HXMT), named "Insight", is the first X-ray astronomy satellite of China launched in 2017 June. The three main payloads onboard \textit{Insight}-HXMT, which are the Low Energy X-ray telescope (LE), the Medium Energy X-ray telescope (ME) and the High Energy X-ray telescope (HE), cover a wide energy bands. The geometrical area of LE is 384 $\rm cm^2$ covering 1-15 keV, ME has a total area of 952 $\rm cm^2$ with an energy range of 5-30 keV, and HE has a total geometrical are of about 5100 $\rm cm^2$ covering 20-250 keV \citep{Zhangsn2020}.

\textit{Insight}-HXMT has monitored the outburst of MAXI J1535$-$571 around its peak intensity in the period 2017 September 6 to 23. The \textit{Insight}-HXMT Data Analysis Software (HXMTDAS) v2.04 is used to reduce the data with the following criteria: The pointing offset is less than 0.04$\degr$, the elevation angle is set to be at least 10$\degr$, the value of the geomagnetic cutoff rigidity is greater than 8, and events must be detected more than 300 s before and after the South Atlantic Anomaly (SAA) passage. To estimate the light curve and the spectrum background count rate, MEBKGMAP and HEBKGMAP in HXMTDAS are used. All the light curves are generated with 0.0078125 s time resolution within the energy band of 10-35 keV and 26-100 keV for ME and HE respectively. To investigate the energy dependence of timing parameters, we separate the energy bands into the following: 10.0-12.0 keV, 12.0-14.0 keV, 14.0-17.0 keV, and 17.0-35.0 keV for ME; 26.0-30.0 keV, 30.3-35.5 keV, 35.5-55.0 keV, and 55.0-100.0 keV for HE.

Wavelet analysis is used to study the time-frequency space information of the light curves \citep[for the technique details, see][]{Chen2022,Torrence1998}. The Morlet wavelet with $m = 6$ is chosen as the 'mother' wavelet. The univariate lag-1 autoregressive [AR(1)] is used for red-noise calculation, to determine the 95 percent confidence level. All light curves are separated by the good time intervals (GTIs) before performing wavelet method because smooth, continuous time series make the results more reliable.

To study the short lasted quasi-periodic oscillations (QPOs), Power Density Spectra (PDS) are produced with only 8s data intervals. Fractional rms-squared normalization \citep{Belloni1990,Miyamoto1991} is used after subtracting the Poisson noise. XSPEC v12.12.0 \citep{Arnaud1996} is then used to fit the PDS with a multi-Lorentzian function in the frequency range between 0.1 Hz to 32 Hz.

Fourier cross spectrum is used to calculate the phase-lag \citep{Nowak1999,Uttley2014} with 10.0-12.0 keV regarded as the reference band. The frequency range used to calculate the average phase-lag is $\nu_0 \pm FWHM/2$, where $\nu_0$ is the QPO centroid frequency, and $FWHM$ is its full width at half-maximum.

For the QPO and non-QPO spectra, we still fit with XSPEC v12.12.0. Because only ME and HE is included in our data, 10-27 keV (ME) and 27-100 keV (HE) are adopted for the fitting with no systematic error added.

\section{Light Curve Analysis}
\subsection{Full energy bands}
\label{sec:3.1}
\begin{figure*}
	\includegraphics[width=1.3\columnwidth]{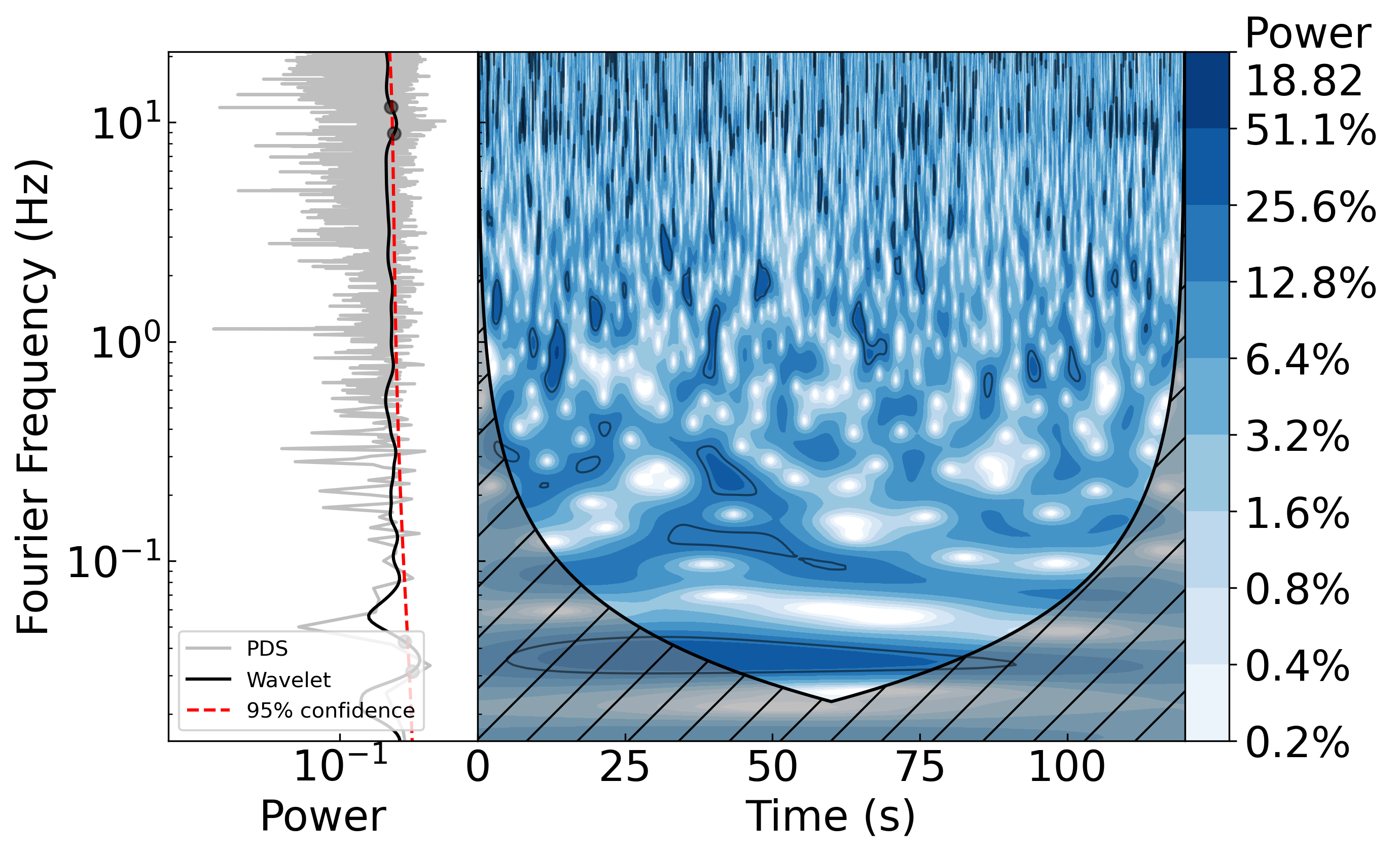}
	\caption{The wavelet result of the first GTI of 701 ME (10-35 keV). PDS (grey line), global wavelet spectrum (black line), and 95\% confidence level (red line) are plotted in the left panel. The contour plot (local wavelet spectrum) are shown in the right panel. Regions with greater than 95\% confidence level are circled with black lines in the contour plot, and the cone of influence area is marked with gray hashed lines. Color bars of the contour plot is presented on the right side. The value (i.e. 18.82) on the top shows the maximum power in the contour plot, and the color scale represent the percentage of that maximum power.}
	\label{fig:701ME00}
\end{figure*}

\begin{figure*}
\begin{minipage}{0.48\textwidth}
\includegraphics[width=1\textwidth]{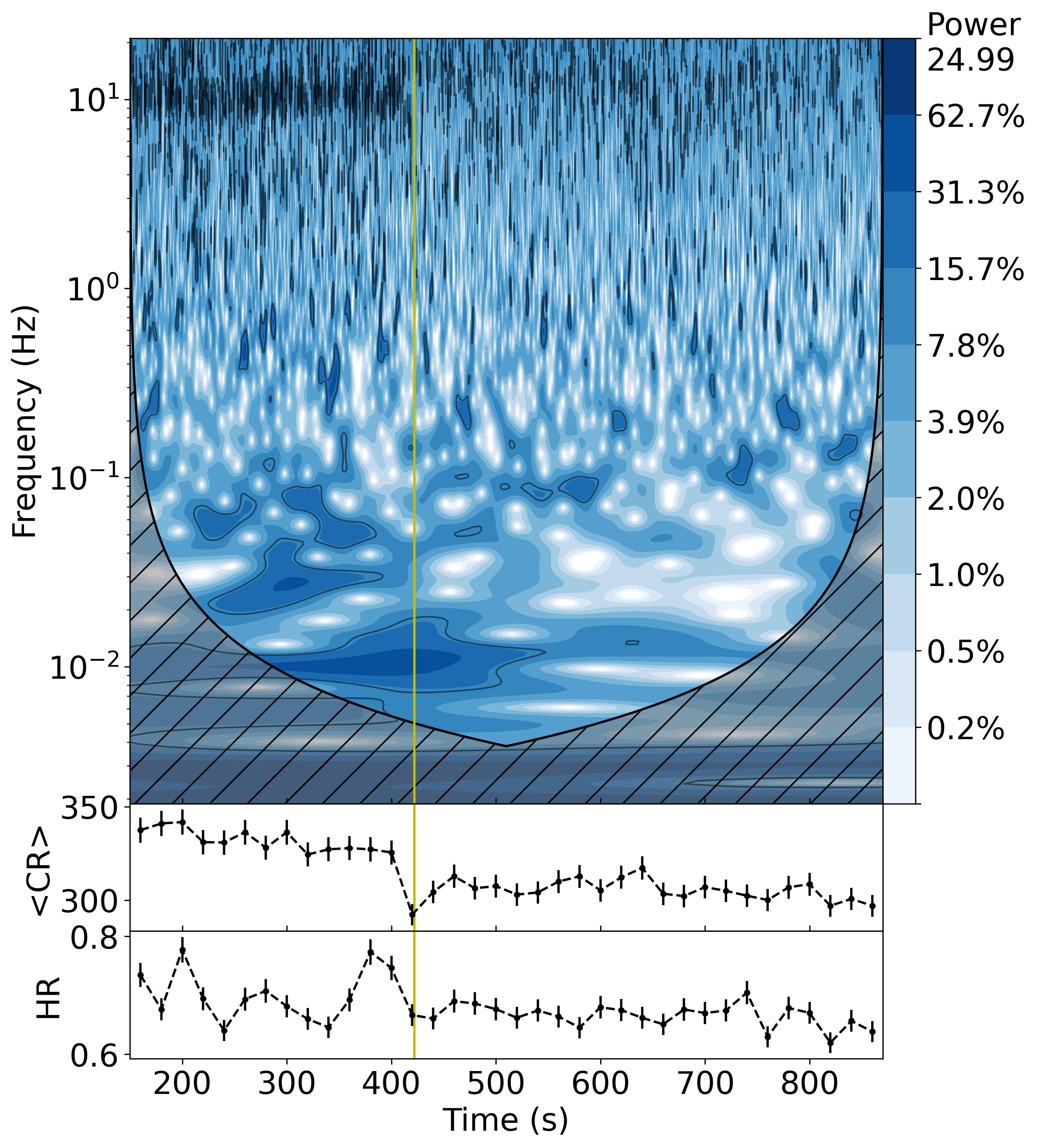}
\end{minipage}
\begin{minipage}{0.48\textwidth}
\includegraphics[width=1\textwidth]{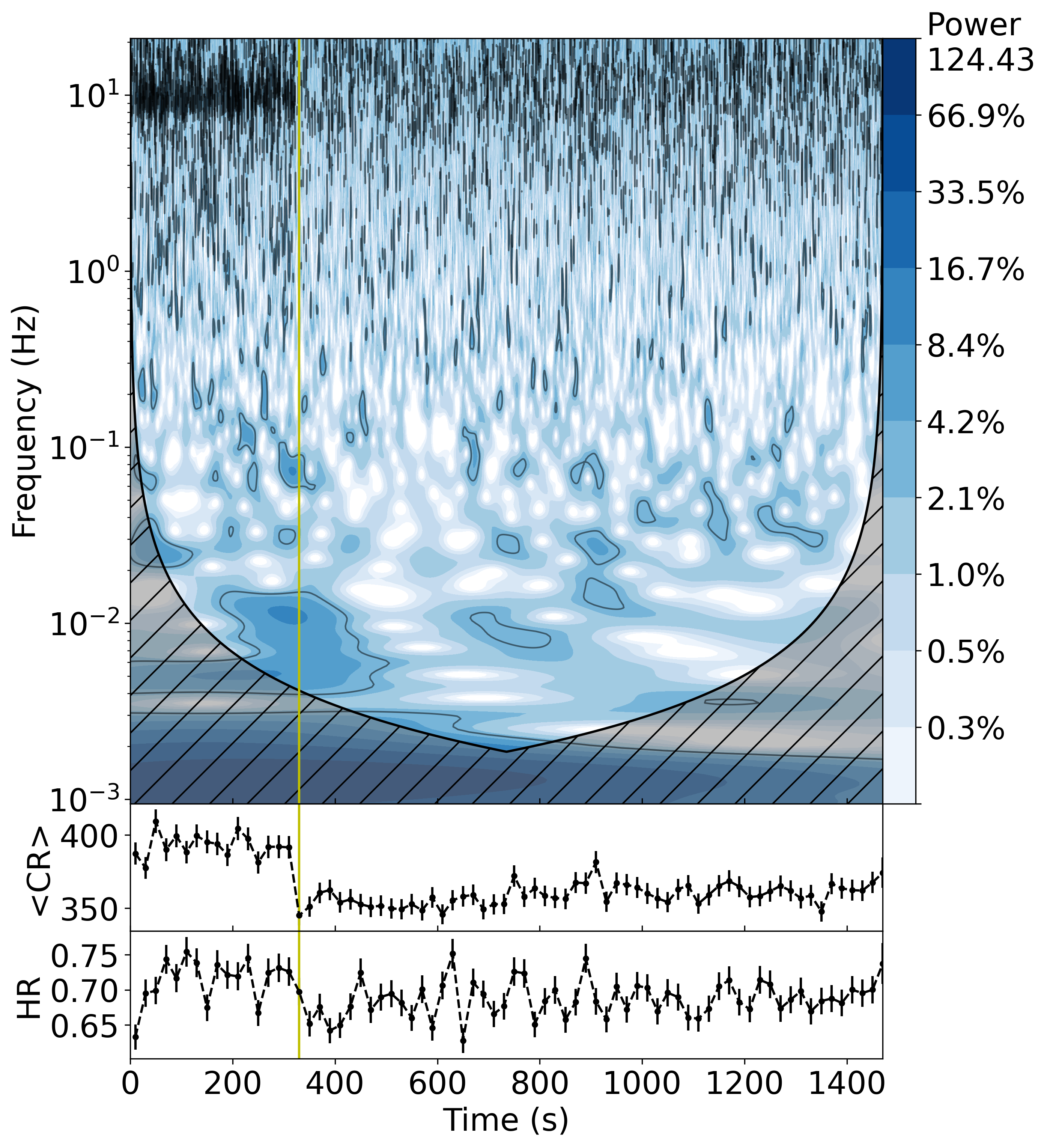}
\end{minipage}
\caption{The wavelet results, <CRs> and HRs of 701 (left) and 904 (right) over time. The local wavelet spectra are presented in the top panels, with the same elements as described in Figure~\ref{fig:701ME00}. <CRs> of every 20 s are shown in the middle panels. The local wavelet spectra and <CRs> are both the results of ME (10-35 keV). In the bottom panels, we show the HR of 12-14 keV/10-12 keV averaged in every 20 s. The yellow lines are drawn in all the panels to distinguish the QPO and non-QPO areas.}
\label{fig:wave}
\end{figure*}

Because wavelet results are separated by the GTIs, it is necessary to briefly introduce the distribution of our GTIs in the Observation IDs P011453500701 (hereafter 701) and P011453500904 (hereafter 904). For the ME of 701, the first GTI is about 120 s, containing relative weak QPO signals as shown in Figure~\ref{fig:701ME00}. Then the second GTI starts from $\sim$150 s to $\sim$900 s, and strong QPO lasts from 150 -- 422 seconds and then disappears after that as presented in Fig~\ref{fig:wave}. The third GTI begins at $\sim$3000 s, but no QPO signal is detected since then. For the HE of 701, the first GTI ends at $\sim$140 s and starts at $\sim$200 s for the second GTI. For the ME of 904, only the first GTI has QPO signals detected within the early $\sim$330 s (see Fig~\ref{fig:wave}), and the HE has the similar GTI (also see Table~\ref{tab:hrcr}). The second GTI of 904 starts from $\sim 3000$ s, and no QPO is detected.

In Figure~\ref{fig:701ME00}, the wavelet spectrum for the first GTI of 701 is presented. The left panel is the power density spectrum and the global wavelet spectrum showing the weak QPO with the frequency at $\sim 9.9$ Hz, while the local wavelet spectrum in the right panel reflects the QPO signal evolution with time in a high time resolution. The QPO signal distributes sparsely, some time disappears for several to tens of seconds.

For the second GTI of 701 and the first one of 904, the wavelet results are shown in Fig~\ref{fig:wave}. For both observations, the QPOs appeared only in the early time intervals and then disappeared. We have presented the mean count rates (<CRs>) and hardness ratios (HRs) averaged in every 20 seconds in Fig~\ref{fig:wave}, where HR is the count ratio of 12-14 keV/10-12 keV. A sudden decrease can be noted in the <CR> evolution, while a weak decline can be noticed in the HR plots when QPO disappears, but the change is not obvious. To check the differences further, we average the QPO/non-QPO time window within each GTI to calculate <CRs> and HRs. The results are shown in Table~\ref{tab:hrcr}. Because LE bands are missing for both of the observations, we use hardness ratios defined as HR1$=(12-14){\rm keV}/(10-12){\rm keV}$  and HR2$=(30.3-35.5){\rm keV}/(26-30.3){\rm keV}$. For both 701 and 904, the <CRs> and HRs of the QPO regime are always larger than the non-QPO regime, respectively.

To justify the QPO properties in wavelet analysis, we define S-factor, which is related to the local wavelet spectrum, as
\begin{equation}
    S = \frac{\tau_{eff}}{\tau_{sel}},
\end{equation}
where $\tau_{eff}$ and $\tau_{sel}$ are the effective oscillation time duration and the total lengths, respectively, of the selected time range(s) in seconds. The effective oscillation time is calculated as the following. Firstly, We define the effective QPO frequency range as the intersection range between the 95 percent confidence level and the global wavelet spectrum containing the centroid frequency. Then if any power in the local wavelet spectrum within this frequency range is greater than the 95 percent confidence level, the corresponding time is selected as the effective oscillation time. Meanwhile, the error of the S-factor is calculated based on the telescope time resolutions as shown in Table~\ref{tab:hrcr}.

For the observations in 701, the QPO signals evolve from the first GTI to the second, the S-factor changes in three epochs. For ME, in the first GTI from 0 -- 120 s, the weak QPO has $S\sim 15\%$, and during the second GTI, from 150 -- 422 s, the QPO is stronger with $S\sim 23\%$, then after 422 s, the QPO disappears, $S\lesssim 7\%$.

\begin{figure*}
\begin{minipage}{0.48\textwidth}
\includegraphics[width=1\textwidth]{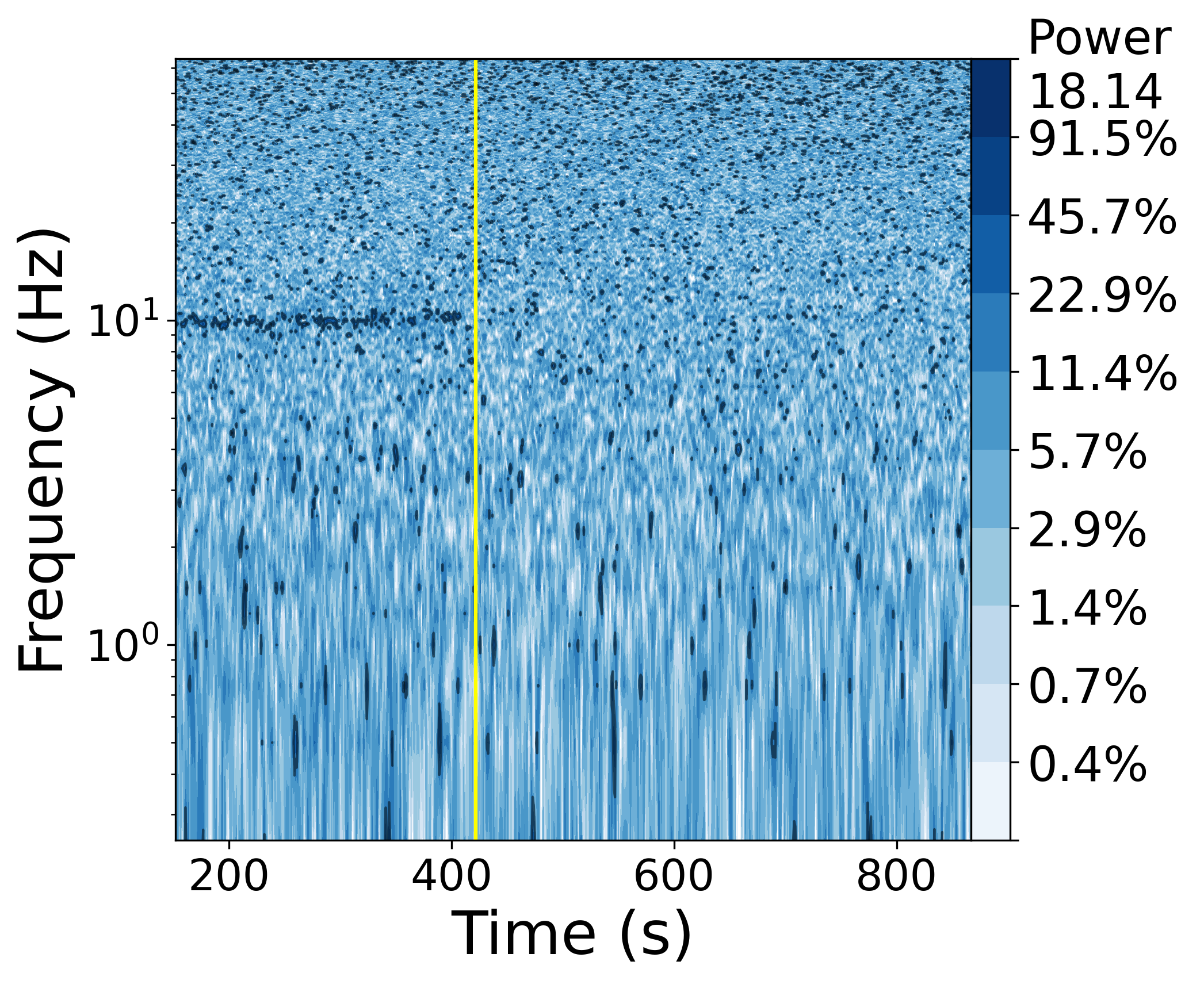}
\end{minipage}
\begin{minipage}{0.48\textwidth}
\includegraphics[width=1\textwidth]{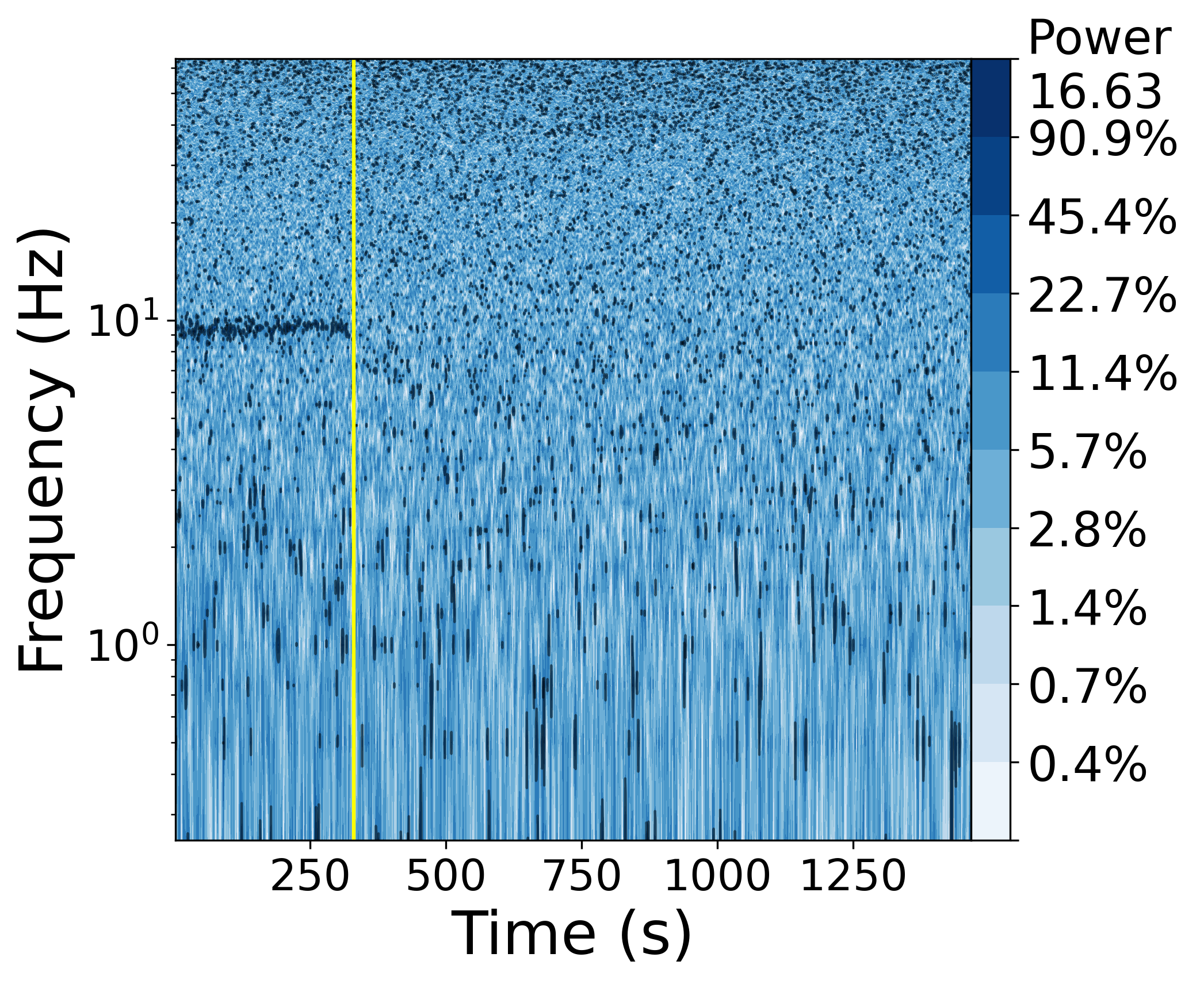}
\end{minipage}
\caption{The sliding dynamical PDS results of 701 (left) and 904 (right) over time. The elements in the plot are the same as those described in the contour plot of Figure~\ref{fig:701ME00}.}
\label{fig:dynamicalPDS}
\end{figure*}

To support our results, a more conventional approach is used to make a comparison. We computed PDS throughout one GTI using overlapping 4 s intervals with a new interval defined every 1 s. To compare with the wavelet result, we normalize the light curve and power in the same way as the wavelet process, without Poisson noise subtracted. We also use the [AR(1)] and the 95 percent confidence level to define the significance of the signal, and the results are presented in Fig~\ref{fig:dynamicalPDS}. Based on our sliding method, the significance of each second is jointly determined by 4 intervals. If at least 3 intervals are determined as significant, then this second will be classified into $\tau_{eff}$ of the S-factor. Similarly, the lower and upper limit of the S-factor are calculated if all or at least 1 interval is significant, respectively. For consistency, we use the same effective QPO frequency range as that used in wavelet S-factor calculation of the corresponding observation. As a result, the S-factor in the QPO regime is $\sim 0.75_{-0.25}^{+0.20}$, and $\sim 0.15_{-0.10}^{+0.50}$ in the non-QPO regime. These values are larger compared to the wavelet calculated S-factor, especially for the QPO regime, the error bars are much larger than the wavelet results since the time resolution of PDS is much lower.

\begin{table*}
    \scriptsize
	\centering
	\caption{The mean count rates and hardness ratios of 701 and 904. }
	\label{tab:hrcr}
	\begin{tabular}{ccccccccc}
		\hline
		Obs IDs & MJD & QPO type & ME time range$^{\ddag}$ & ME <CR> & HR1$^*$ & HE time range$^{\ddag}$ & HE <CR> & HR2$^*$ \\
		\hline
		701 & 58015.974 & weak QPO & 1$^{\dag}$: 0-120s & $343.76 \pm 2.77$ & $0.705 \pm 0.008$&1$^{\dag}$: 0-143s & $341.33 \pm 2.76$ & $0.639 \pm 0.007$ \\
		   & & strong QPO& 2$^{\dag}$: 150-422s& $330.18 \pm 1.77$ & $0.700 \pm 0.005$ & 2$^{\dag}$: 199-422s&$337.41 \pm 2.00$ & $0.634 \pm 0.005$\\
		   & & non-QPO&2$^{\dag}$: 422-870s & $305.48 \pm 1.28$ & $0.665 \pm 0.004$ & 2$^{\dag}$: 422-893s&$258.95 \pm 1.08$ & $0.598 \pm 0.004$\\
		904 & 58017.529 & QPO& 1$^{\dag}$: 0-330s & $391.53 \pm 1.91$ & $0.713 \pm 0.005$ & 1$^{\dag}$: 0-330s&$416.80 \pm 2.03$ & $0.646 \pm 0.004$\\
		   & & non-QPO&1$^{\dag}$: 330-1470s & $359.01 \pm 0.94$ & $0.686 \pm 0.003$ & 1$^{\dag}$: 330-1494s&$320.72 \pm 0.84$ & $0.636 \pm 0.002$\\
		\hline
		\multicolumn{9}{|l|}{$^*$ HR1: 12-14 keV/10-12 keV; HR2: 30.3-35.5 keV/26-30.3 keV.} \\
		\multicolumn{9}{|l|}{$^\dag$ 1: the first GTI; 2: the second GTI} \\
		\multicolumn{9}{|l|}{$^\ddag$ ME time resolution: 240 µs; HE time resolution: 4 µs} \\
		
	\end{tabular}
\end{table*}

\begin{figure*}
\begin{minipage}{0.48\textwidth}
\includegraphics[width=1\textwidth]{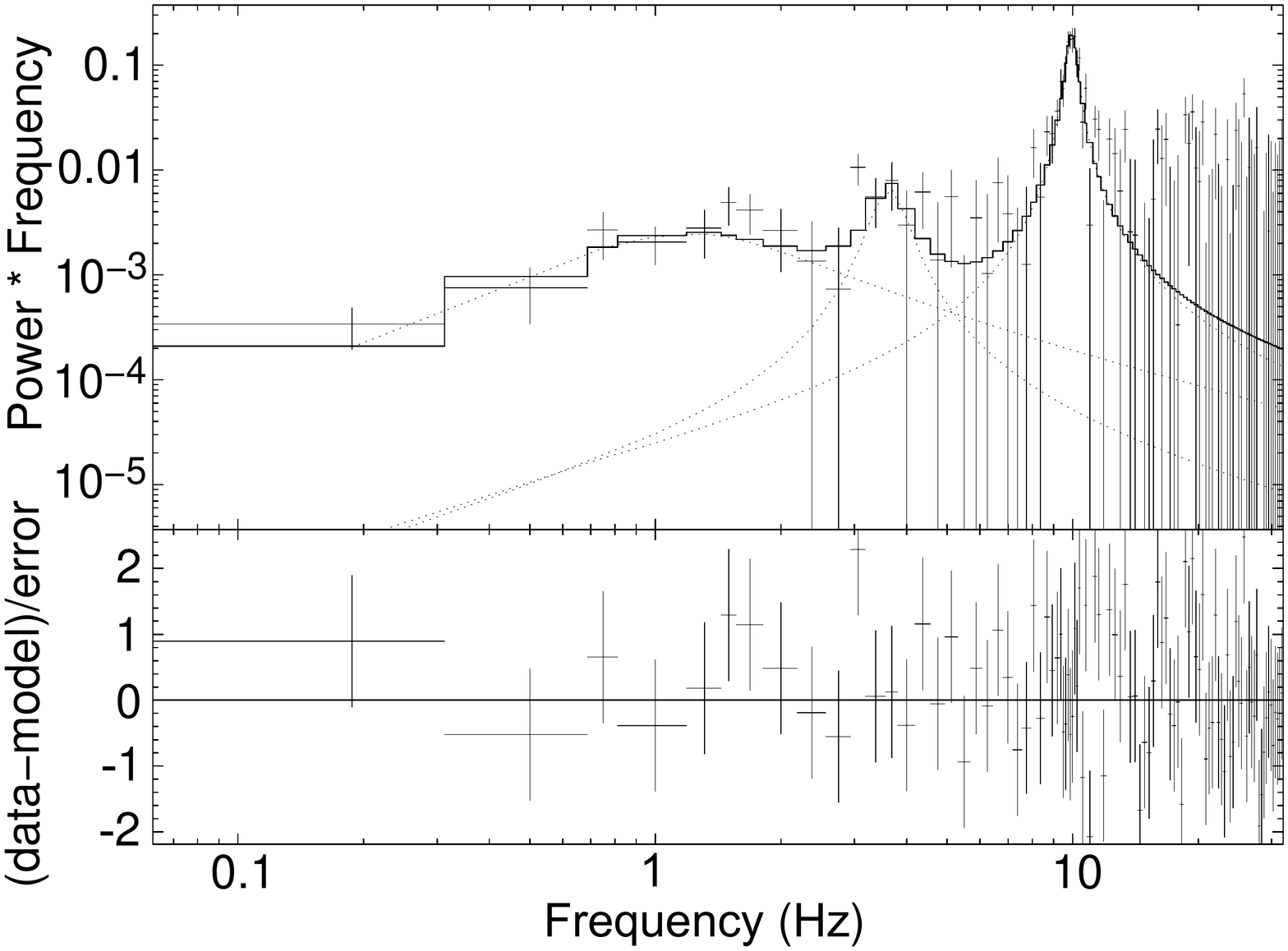}
\end{minipage}
\begin{minipage}{0.48\textwidth}
\includegraphics[width=1\textwidth]{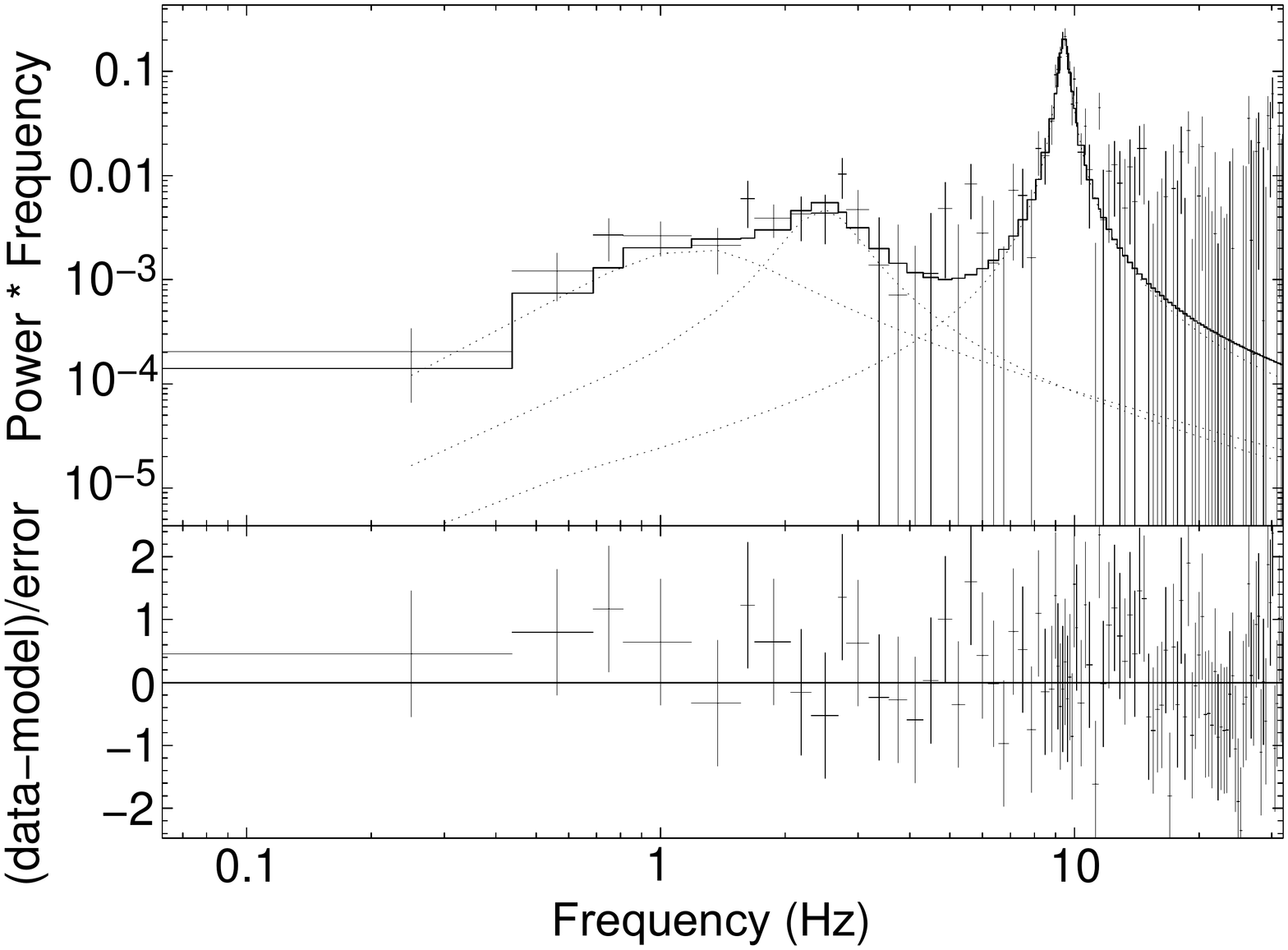}
\end{minipage}
\caption{The PDS of 701 (first 422s, including the 1st and part of the 2nd GTI) and 904 (first 330s) using the 10-35 keV ME data. Left panel: The PDS (top) and the residuals (bottom) of 701. Right panel: The PDS (top) and the residuals (bottom) of 904.}
\label{fig:pds}
\end{figure*}

\begin{figure}
	\includegraphics[width=\columnwidth]{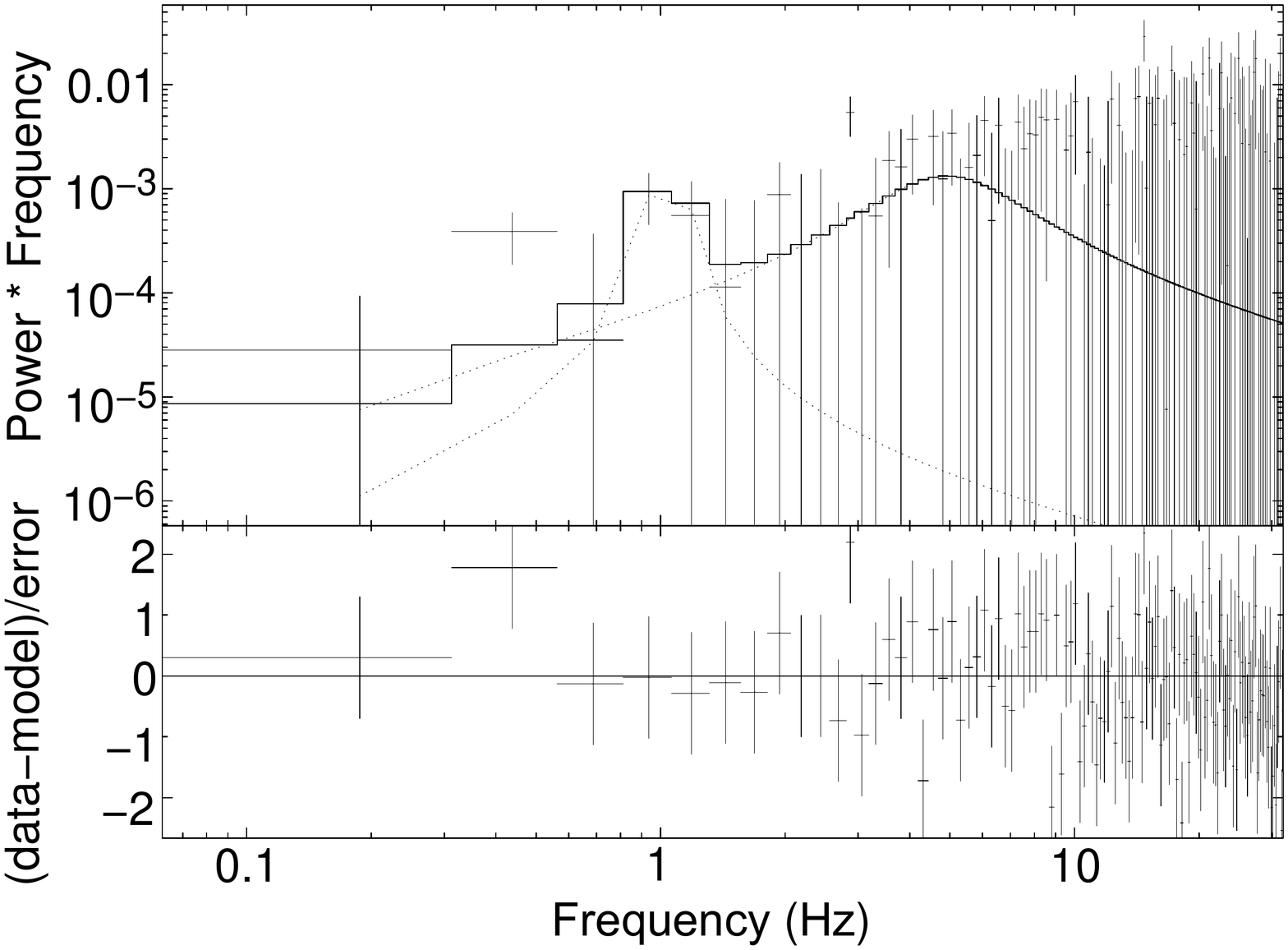}
	\caption{The PDS of the non-QPO part in the 1st GTI of 904 ($\sim 330 - 1470 s$, 10-35 keV). The model is the same as that used in Fig~\ref{fig:pds}, with the frequency and the width of the QPO Lorentzian fixed to the corresponding values in Fig~\ref{fig:pds}. The normalization of the QPO component is quite weak ($\sim 1.8\times 10^{-8}$), thus cannot be seen in the plot.}
	\label{fig:nonQPO PDS}
\end{figure}

In addition, the average PDSs are computed with the light curves of the first 422 s (including the 1st GTI and QPO part of the 2nd GTI) of 701 and the first 330 s of 904, respectively. The PDS examples fitted with multi-Lorentzian functions are shown in Fig~\ref{fig:pds}, and the QPO fitting results are also shown in Table~\ref{tab:qpo}. The QPO properties in both 701 and 904 are very similar to each other, with their centroid frequency $\sim$10 Hz, FWHM $\sim$0.6 Hz, and rms amplitude around 14\%. Then the quality factor of these QPOs is generally around 16. We also use the same multi-Lorentzian model to fit the non-QPO part of the 2nd GTI of 701 and the 1st GTI of 904, for both ME and HE. However, the QPO components are really weak in the non-QPO regimes, thus we fix the frequency and the width of the QPO Lorentzian to the values in the corresponding QPO Lorentzian in the QPO regime. The QPO rms values are really weak for all the fittings in the non-QPO regime, with maximum value of rms $\lesssim 1\%$. A non-QPO example of the fitting is shown in Fig~\ref{fig:nonQPO PDS}.

\begin{table}
    \scriptsize
	\centering
	\caption{QPO properties of 701 and 904 using the same time segments as Figure~\ref{fig:pds}. }
	\label{tab:qpo}
	\begin{tabular}{ccccc}
		\hline
		Obs IDs & Energy band & $\nu_0$ (Hz) & FWHM (Hz) & rms (\%) \\
		\hline
		701 & ME (10-35 keV) & $9.92 \pm 0.06$ & $0.63 \pm 0.12$ & $14.1 \pm 0.9$\\
		   & HE (26-100 keV) & $10.02 \pm 0.05$ & $0.56 \pm 0.14$ & $14.1 \pm 1.0$\\
		904 & ME (10-35 keV) & $9.43 \pm 0.05$ & $0.55 \pm 0.11$ & $14.1 \pm 0.9$ \\
		   & HE (26-100 keV) & $9.46 \pm 0.05$ & $0.66 \pm 0.15$ & $15.3 \pm 0.9$\\
		\hline
	\end{tabular}
\end{table}

\subsection{Energy separated bands}
\begin{figure}
	\includegraphics[width=\columnwidth]{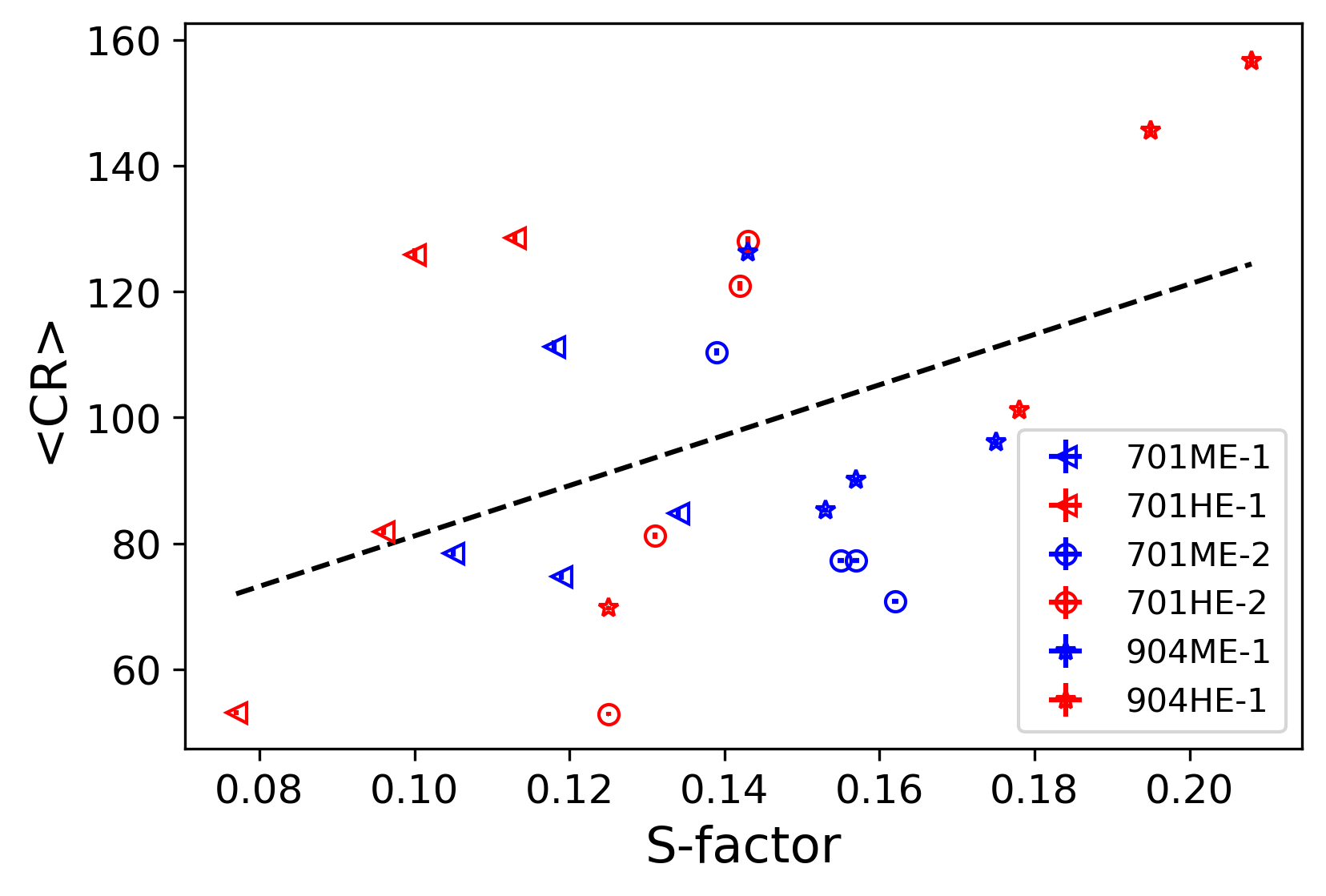}
	\caption{The relation of S-factor and <CR> in different energy bands in the QPO regime. Left triangle, circle and star refer to the 1st GTI of 701, the 2nd GTI of 701 and the 1st GTI of 904, respectively. Blue and red represent ME and HE, respectively. The whole data are fitted with a linear function presented with a black dashed line. The error bars are also plotted, but are barely seen in the plot since they are quite small.}
	\label{fig:S-CR}
\end{figure}

As mentioned earlier, ME and HE are both separated into four energy bands to study the energy dependence of the light curve parameters. The wavelet analysis is applied to the lightcurves for the different energy bands, so that we can estimate the S-factors and mean count rates in eight energy bands for three QPO regimes. We show the relation of <CR> and S-factor in Figure~\ref{fig:S-CR}. Meanwhile, we use a linear function to fit these data points. A Spearman test shows that the Spearman rank correlation $r=0.30$ and the corresponding p (or probability) is 0.15. Normally, a p-value of 0.05 (5\%) or less is typically considered statistically significant. Thus the points show a weak correlation. Moreover, we run the Spearman test on the HE data (red) only, and a strong positive correlation with $r=0.57$ and $p=0.05$ is shown, while the test on the ME data (blue) only shows $r=-0.15$ with $p=0.63$. These tests showed us that, first of all, the S-factor is possibly related to the mean count rate, and secondly, HE and ME may show different correlations.

\begin{figure}
	\includegraphics[width=\columnwidth]{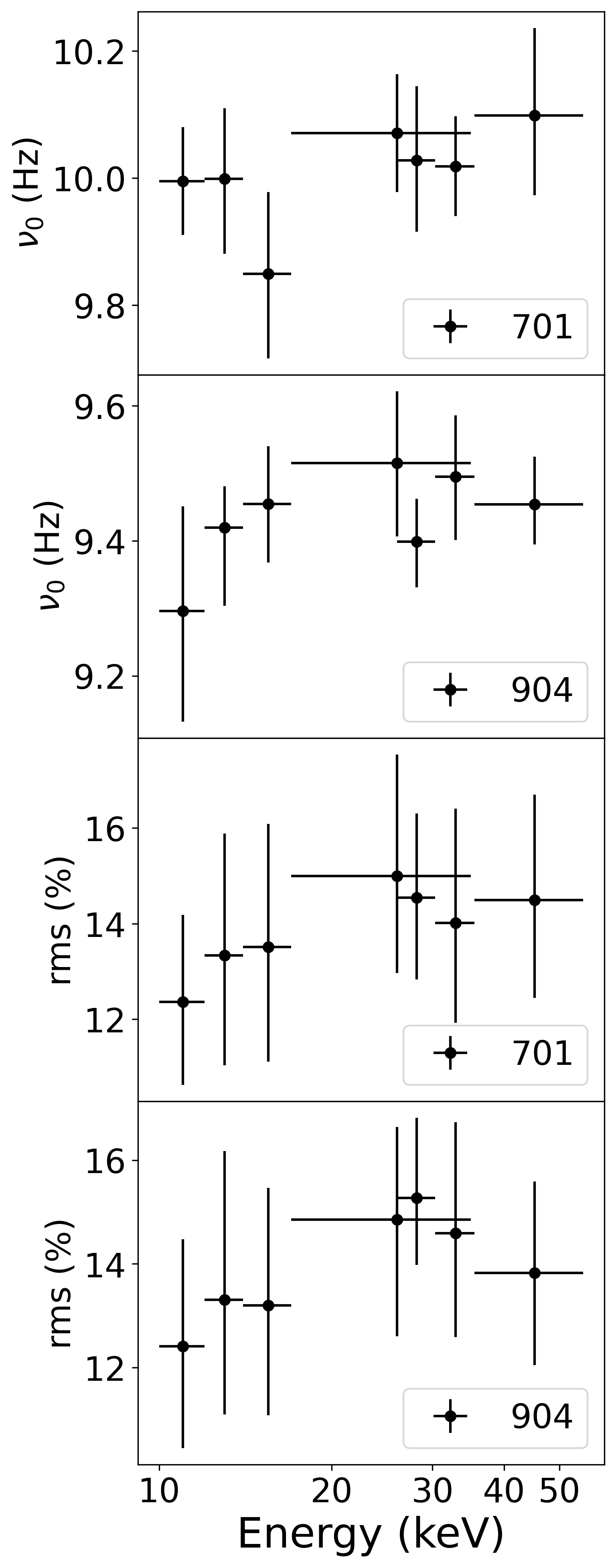}
	\caption{The centroid frequency and rms amplitude over different energy bands.}
	\label{fig:QPOs}
\end{figure}

The separated energy bands are also used to study the QPO properties. The time ranges used for PDS generation are exactly the same as Section~\ref{sec:3.1}, and the results are shown in Figure~\ref{fig:QPOs}. The QPOs in the 55-100 keV band are quite weak with very narrow width (FWHM < 0.1), and its rms is much smaller than the other energy bands with a value of $\sim8.5\%$ \citep[see also][]{Zhang2022}. Therefore, in order to avoid inaccurate fitting, the results of this band are not included in the plot. We do not find significant changes in the properties of QPOs at different energy bands, as the changes are more conspicuous below 10 keV \citep{Huang2018}.

\begin{figure}
    \begin{minipage}{0.5\textwidth}
		\includegraphics[width=1\textwidth]{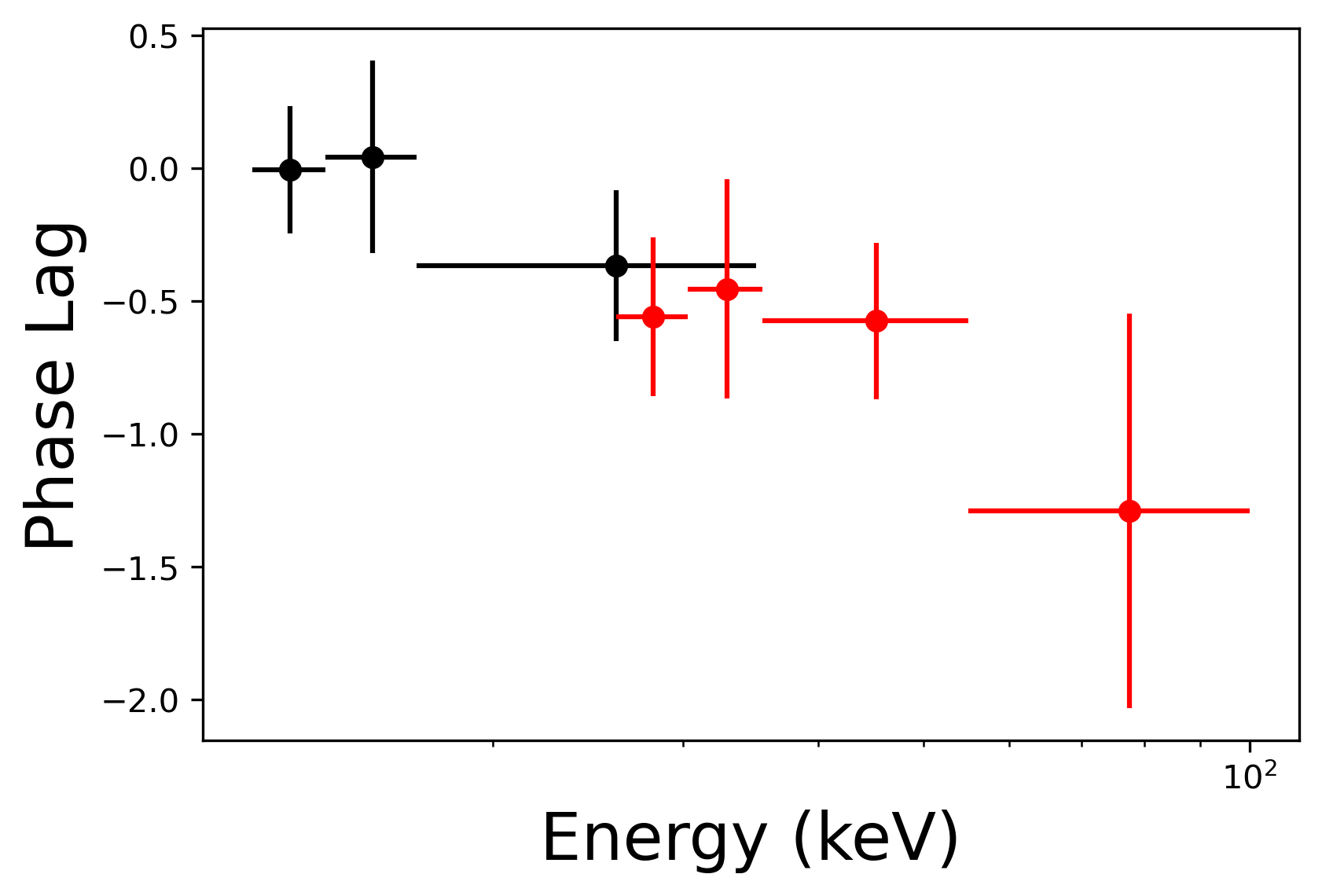}
	\end{minipage}
	\begin{minipage}{0.5\textwidth}
		\includegraphics[width=1\textwidth]{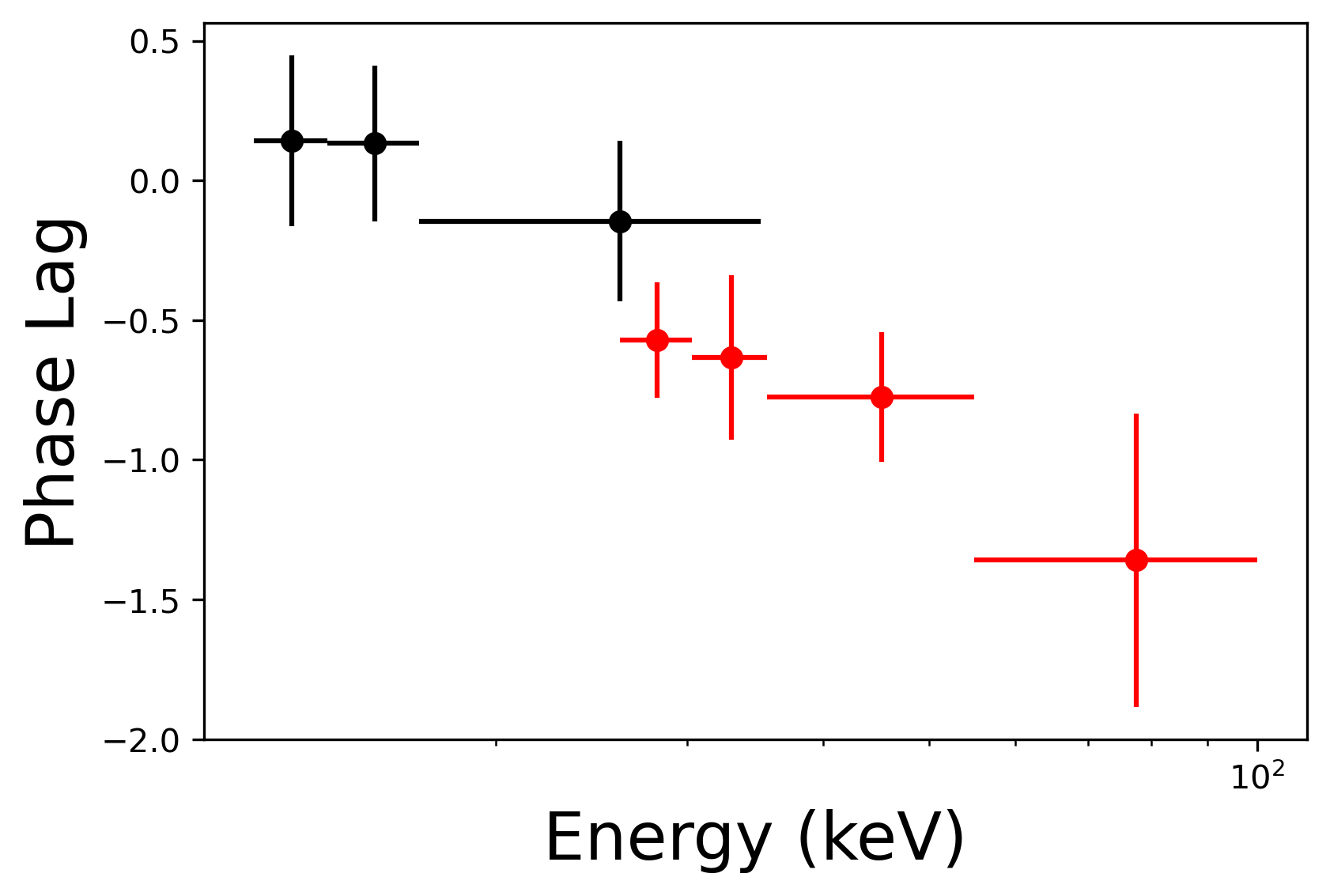}
	\end{minipage}
	\caption{The QPO phase lag spectra of 701 (top panel) and 904 (bottom panel). The black crosses indicate the ME data, while the red crosses are the HE data.}
	\label{fig:phaseLag}
\end{figure}

The phase lag spectra of 701 and 904 are also plotted in Figure~\ref{fig:phaseLag}. The phase lags are negative and the difference increases with energy. The obvious soft lags combined with the high QPO frequencies support the previously reported high inclination angle of this source \citep{Eijnden2017}.

\section{Spectral Analysis}

\begin{figure*}
\begin{minipage}{0.48\textwidth}
\includegraphics[width=1\textwidth]{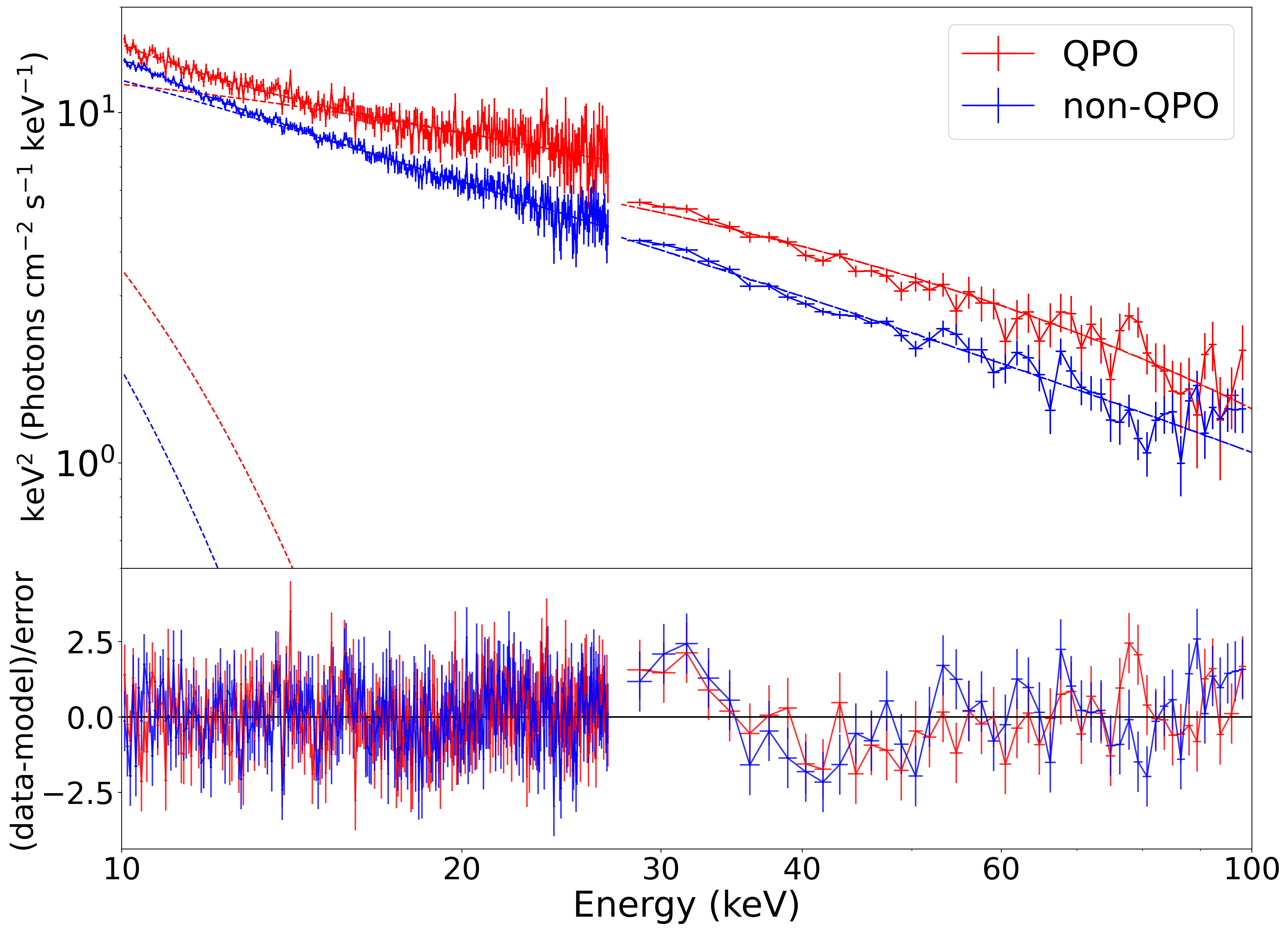}
\end{minipage}
\begin{minipage}{0.48\textwidth}
\includegraphics[width=1\textwidth]{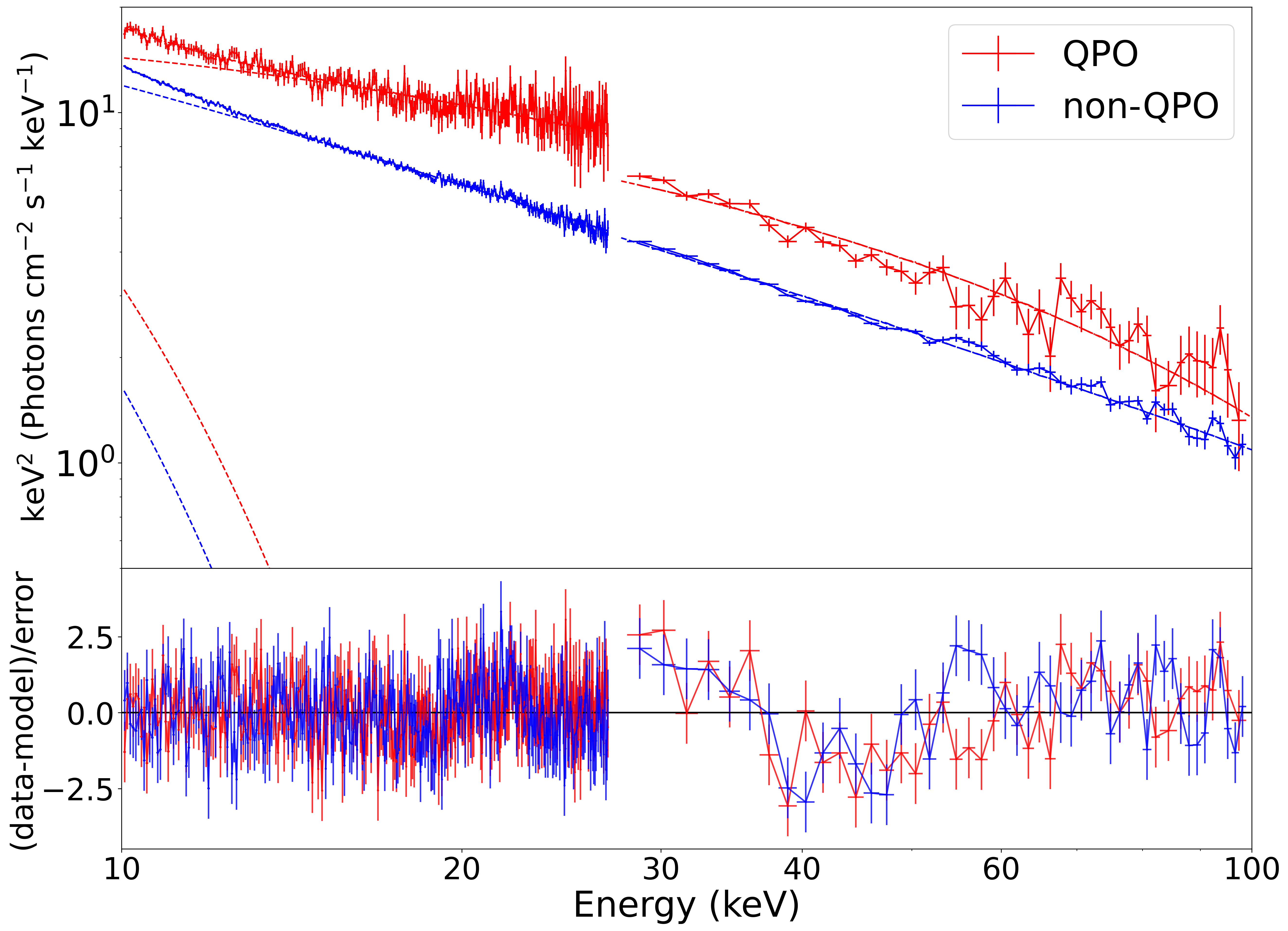}
\end{minipage}
\caption{The QPO and non-QPO spectra fitting results of 701 and 904 using the energy band of 10-100 keV. The QPO spectrum of 701 contains the first GTI and the QPO segment of the second GTI, the QPO spectrum of 904 contains the QPO segment of the first GTI, while the non-QPO spectra of 701 and 904 contain the whole remaining time of that observation. The red color refer to the QPO spectra, while the blue color represent the non-QPO spectra. Left panel: The fitting results of 701. The spectra and model components are shown on the top, and the residuals are shown on the bottom. Right panel: The spectra, model components and residuals of 904.}
\label{fig:spec}
\end{figure*}

To study the spectral differences between the QPO time and the non-QPO time, we separate the time-averaged spectra into QPO spectra and non-QPO spectra. The spectra are all regrouped to ensure that at least 30 counts are included in each energy bin. Because only ME and HE are included in 701 and 904, we start by considering $const*tbabs*cutoffpl$ as the fitting model. However, residuals are quite clear below 15 keV, suggesting that a disc component still exists. To compare with the results in \cite{Chen2022}, we choose $const*tbabs*(diskbb+cutoffpl)$ as our final model, where the parameter $N_H$ in the $tbabs$ is fixed at $3.8\times 10^{22} \rm \ cm^{-2}$. The fitting results are listed in Table~\ref{tab:fitting} and Figure~\ref{fig:spec}. Because the normalization of the $cutoffpl$ model in XSPEC is the flux density of the model at 1 keV, which is out of the data, we use $cflux$ in XSPEC in 10 - 100 keV to calculate the flux of the component, and the flux is shown in Table~\ref{tab:fitting} instead of the normalization. In the non-QPO spectrum fitting, the values of $E_{cut}$ always reach the upper limit of 500 keV, so the positive error is not given.

\begin{table*}
	\centering
	\caption{Spectral fitting results of 701 and 904. The model used is $const*tbabs*(diskbb+cutoffpl)$, with $N_{H}$ fixed at $3.8\times 10^{22} \rm \ cm^{-2}$.}
	\label{tab:fitting}
	\begin{tabular}{ccccccc}
		\hline
		Obs	&	 $T_{in}$	 &	 diskbb flux$^{\dag}$	 &	 $\Gamma$	 &	 $E_{cut}$	 &	 cutoffpl flux$^{\dag}$	 &	 $\chi^2/d.o.f.$  \\
		     & (keV) & ($10^{-10} erg/cm^2/s$) & & (keV) & ($10^{-10} erg/cm^2/s$) & \\
		\hline
		\multicolumn{7}{|c|}{\bf{QPO spectrum}} \\
		701	&$	1.42\pm 0.21 	$&$	11.04 \pm 4.56 $&$	2.32 \pm 0.14 	$&$	77.7 \pm21.7 $&$	234.96 \pm3.51$&	349.68 	/	332	\\

904	&$	1.30 \pm 0.23 	$&$ 8.89\pm 4.26$&$	2.26\pm 0.12	$& $58.9\pm 11.6$&$277.33\pm 3.51$&	378.01 	/	332	\\

		\multicolumn{7}{|c|}{\bf{non-QPO spectrum}} \\
		701	&$	1.15 \pm 0.14 	$&$	4.29 \pm 1.37$&$	2.99 	\pm	0.04 	$&	$500_{-209.1}^{\quad-}$ 	&$	174.58 \pm 1.21 $&	377.33 	/	332	\\

904	&$	1.16 \pm 0.06 	$&$	3.97 \pm 0.49$&$	2.97 	\pm	0.01 	$&	$500_{-76.37}^{\quad-}$ 	&$	170.61 	\pm 0.39$&	403.65 	/	332	\\

		\hline
		\multicolumn{7}{|l|}{$^\dag$: 10 - 100 keV} \\
	\end{tabular}
\end{table*}

Figure~\ref{fig:spec} shows that the spectral difference in 904 is more obvious then that in 701. At around 10 keV, the corona components of QPO and non-QPO spectrum in both 701 and 904 are very close to each other, but the difference gradually becomes significant as the energy increases. The disc components of the QPO spectra show relatively higher temperature and higher flux compared to the non-QPO spectra as listed in Table~\ref{tab:fitting}. However, because the LE bands are missing for both 701 and 904, the error bars are quite large. Thus we will not overinterpret this component. The corona components show clear discrepancies. The $\Gamma$ and the cutoff energy in the QPO regime are both much smaller than the non-QPO regime, indicating that the spectra become softer and the corona cools down after the QPO disappears. The corona flux decreases from QPO to non-QPO regime.

\section{Discussion}
\label{sec:discussion}
We have presented the properties of two transient QPOs in two observations of MAXI J1535$-$571 during its outburst in 2017 with the \textit{Insight}-HXMT data. Up to now, eleven observations obtained by the \textit{Insight}-HXMT reported QPOs \citep[i.e. 144-904, see][]{Chen2022,Kong2020} including two transient QPO observations. We have used both wavelet analysis and PDS to study the light curves in different energy bands for these 11 observations (this work and \citealt{Chen2022,Huang2018}). Combined with these analysis results, we try to study both the similarities and differences between the transient QPOs and other QPO signals, then probe the physical properties and origin of the transient QPOs.

\begin{figure}
	\includegraphics[width=\columnwidth]{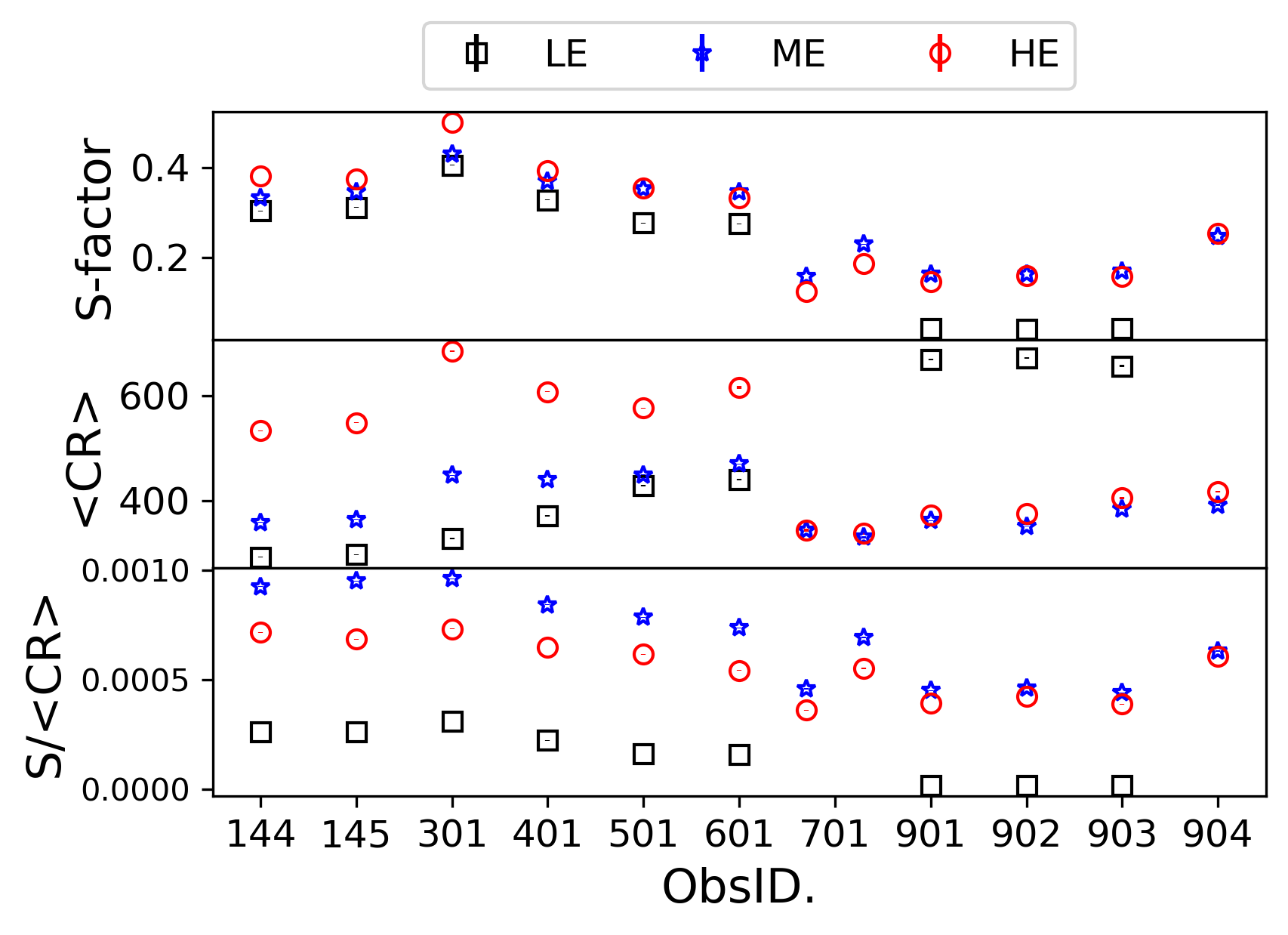}
	\caption{The S-factors (top panel), the mean count rates (middle panel), and the ratios of S/<CR> (bottom panel) of the eleven QPO detected observations in \textit{Insight}-HXMT data. The black square, the blue star, and the red circle are LE, ME and HE data, respectively. The two data points shown in 701 represent the first GTI and the QPO portion of the second GTI, respectively. The data point in 904 only contains the QPO portion of the first GTI.}
	\label{fig:S-CR-obs}
\end{figure}

\begin{figure}
	\includegraphics[width=\columnwidth]{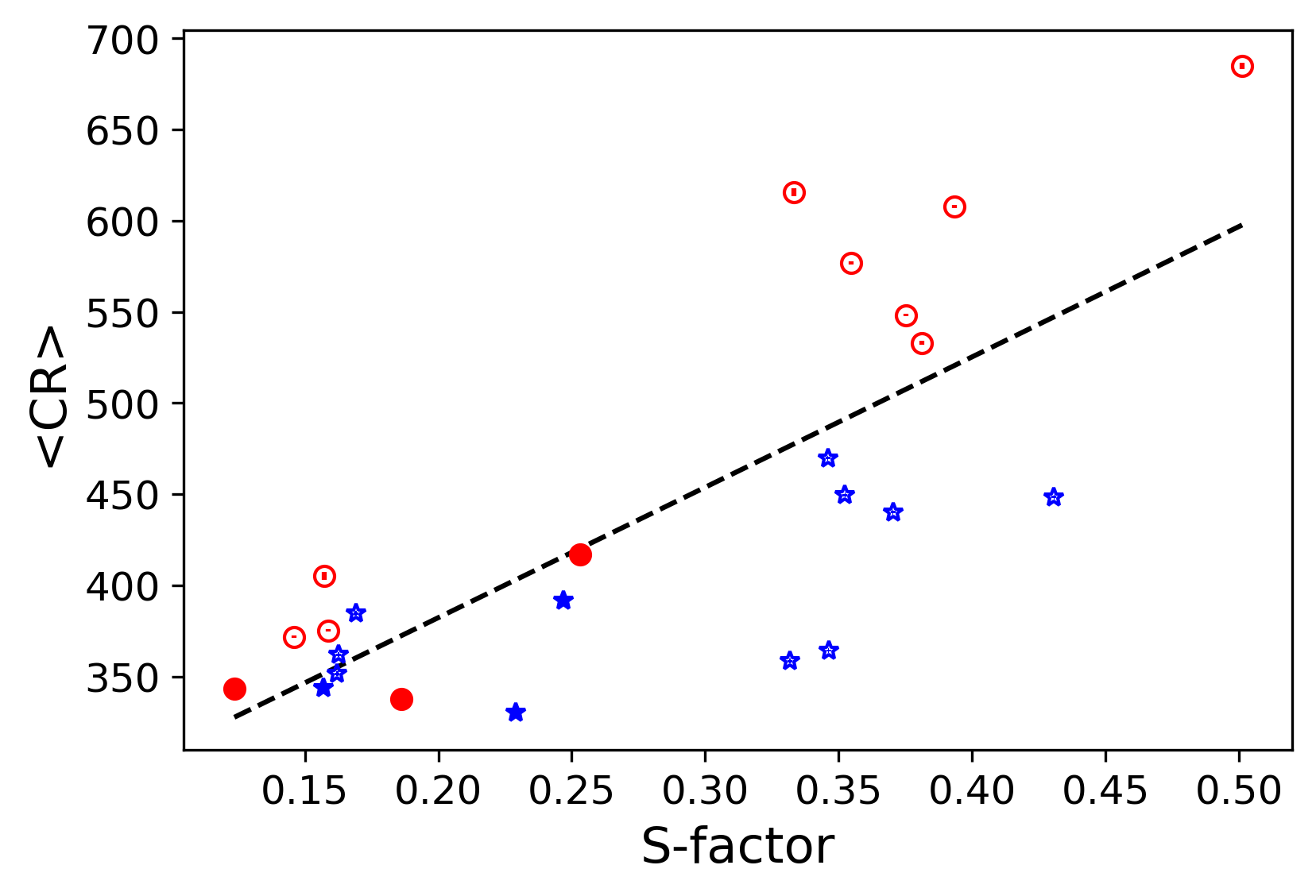}
	\caption{The relation of <CR> and S-factor for ME and HE. The labels represent the same meanings as in Figure~\ref{fig:S-CR-obs}, except that the markers for 701 and 904 have became solid. The 901-903 data are circled with a rectangle, and other data are fitted with a linear function shown as a black dashed line.}
	\label{fig:S-CR2}
\end{figure}

In Figure~\ref{fig:S-CR-obs}, we show the S-factors, the mean count rates, and the ratios between them of the eleven observations with QPO detected. <CRs> and S-factor of ObsID. 144-601 and 901-903 are measured with the same technique as that used in 701 and 904 of this work. Note that the LE count rates are divided by 4 in the middle panel (but remain the original values in the bottom panel) to make the evolution of ME and HE more distinct. Data points of 701 and 904 are taken from this work. The S-factor in 144-601 is about 0.3-0.5, decreases during the first GTI of 701, and then increase to $\sim$ 0.22 during the transient QPO, reaches $\lesssim 0.05$ after the QPO disappearance, and back to $\sim$ 0.16 in 901-903, and finally returns back to $\sim$0.25 in the transient QPO of 904. This demonstrates that the QPO appearance time has become much shorter after 601, or at least the QPO signals have become weaker. In all these observations, the S value of LE is always lower than that of HE, even after an obvious LE flux enhancement as noted in 901-903, indicating that the LE flux increase has barely contribution to the type-C QPO generation. Meanwhile, the sudden decline of the 701 S-factors in ME and HE as their <CRs> decrease, and the gradual increase of both S-factors and <CRs> after that, demonstrating again that S-factor has a positive correlation with <CR> as mentioned earlier. However, as presented in the bottom panel, this correlation is not linear, but more likely to be exponential. There are two eccentric cases in the bottom panel of Figure~\ref{fig:S-CR-obs} showing sudden increase of the $S/<CR>$ ratio, i.e. the second point of 701 and 904. This abrupt change may be related to the discrete jet. Right before the disappearance of a type-C QPO, a precessing hot corona \citep{Ingram2009} may extended in its height due to a transient jet \citep{Sriram2021}. This jet exists for a short time and increases the S-factor as shown in 701 and 904. \cite{Zhang2022} studied the \textit{Insight}-HXMT observations (Obs. 144-501 and 901-903) of the source MAXI J1535$-$571 using the time-dependent Comtonisation model \citep{Karpouzas2020,Bellavita2022}. They found that the corona extended in height in 901-903, and also concluded that the transient jet happens after the corona expansion. However, because 701 and 904 were out of their research, we cannot assert if there is a more abrupt increase in the height of the corona at the beginning of 904. The relation of <CR> and S-factor is plotted in Figure~\ref{fig:S-CR2} with only ME and HE data for all eleven observations. A positive correlation can be noted. A linear function is fitted to show the positive relation, and a Spearman test gives a quite reliable strong positive correlation with $r=0.75$ and $p=2.94\times 10^{-5}$.

\cite{Huang2018} reported that the source went into SIMS during 701 since they suggest the QPO signal in the first $\sim$800s of 701 as a type-B QPO with a centroid frequency of 9.98 Hz and a 5.3\% rms amplitude. A weak red-noise component was also reported around 0.01-1 Hz. The present wavelet analysis based on the updated database shows that the transient QPO in 701 appears in the first $\sim$422s, shorter than that reported in \cite{Huang2018}. This difference may be due to the early HXMT data product and software in the test stage, or the inconsistencies in the GTI criteria we use. The light curves from 422-900 s are taken out separately for wavelet and PDS analysis, but no QPO signal is detected. Secondly, our Lorentz fitting results show that the FWHM of the transient QPO is around 0.6 Hz and rms amplitude is $\sim$14\%, both are very similar to the previous HIMS QPOs (i.e. ObsIDs 144-601, \citealt{Huang2018,Chen2022}). Additionally, the centroid frequency of type-B QPOs is generally around 5-6 Hz or smaller \citep{Motta2011}, and the fractional rms integrated from $1.5-15$ Hz around MJD 58015.974 provided by NICER ($\sim8 \%$) is much larger compared to the rms of the time window when a type-B QPO was detected \citep[$\sim3 \%$;][]{Stevens2018}. Thus, considering the QPO properties of 701 and 904 in our work, these transient QPOs have the similar features to the type-C QPOs in ObsIDs 144-601. Combined with the type-C QPOs detected in ObsIDs 901-903 \citep{Chen2022,Huang2018,Zhang2022}, we suggest that MAXI J1535$-$571 is in the HIMS during ObsIds 144-904.

Although the source is located in HIMS, the transient QPO evolution in 701/904 and spectral differences between ObsIDs 144-601 and 901-903 \citep{Chen2022} may still be caused by the discrete radio jet/flares. \cite{Russell2019} reported the radio activity of MAXI J1535$-$571 during the outburst in 2017 September. They detected a jet knot and suggested the ejection time would be around MJD 58010.8. Two radio flares were also reported. The first one ended on MJD 58017.4 and started after the compact jet quenching ($\sim$MJD 58013.6), and the second one occurred right after the first flare. Because LE data are missing in 701 and 904, we will focus on analyzing the corona components. The spectra of 144-601 and the QPO spectra of 701 and 904 have similar values with $\Gamma$ $\sim$2.3 and $E_{cut} \sim$60 keV, while the spectra of 901-903 and the non-QPO spectra of 701 and 904 show large $\Gamma \sim2.99$ and very high $E_{cut}$ \citep[see][for the spectral results of 144-601 and 901-903]{Chen2022}. Combining the information of X-ray and radio data, the spectral evolution may be related to the radio flares. The disappearance of the QPO in 701 seems likely to be related to the beginning of the first radio flare: the initiation of this radio flare and the softening of the spectrum may occur at the same time. This radio flare continues until MJD 58017.4, which is right between 903 and 904, then the spectrum becomes harder again with $E_{cut} \sim$60 keV (i.e. the QPO time of 904). After a short break, the radio flare starts again, leading to the QPO disappearance and the softening of the spectrum in 904. \cite{Mendez2022} studied the correlation between corona and jet in the source GRS 1915+105. Their results showed that the energy used to accelerate the jet or heat the corona basically has the same origin, i.e. the X-ray corona and the radio jet can morph into each other. As a result, the QPO centroid frequency will decrease when forming jet, and vice versa. This fits very well with the jet launching time inferred by \cite{Russell2019}, since Obs. 301 (MJD $\sim$ 58011.2) shows the lowest QPO centroid frequency. In addition, after the 1st radio flare ends, the QPO centroid frequency increases from $\sim$ 8.5 Hz to $\sim$ 9.4 Hz, indicating that the energy is directed to the corona. Then the energy starts to flow to the radio flux and creates the second flare, leading to the disappearance of the QPO in 904. The QPO disappearance in 701 may be caused by the same mechanism, but more detailed radio data is needed to analyze the changes at that time.

\cite{Sriram2021} studied the transient type-C QPOs in H1743$-$322 and used a broken power-law model to study the hot flow components. They found that $\Gamma_1$ decreased and $\Gamma_2$ increased during the type-C QPO disappearing, where $\Gamma_1$ is the power law photon index for energy less then the break energy $E_{break}$, and $\Gamma_2$ is the power law photon index for energy greater than $E_{break}$, and suggested that $\Gamma_1$ and $\Gamma_2$ might be linked to the outer and inner hot flow, respectively. They hypothesized that the type-C QPO was associated with the base of a jet, and the partial ejection caused the disappearance of the QPO. This interpretation is consistent with our scenario in which the disappearance of QPO in 701 is related to the discrete jet reported in \cite{Russell2019}, along with the partial ejection of the corona. Then an X-ray flare in ME and HE may indicate the replenishment of the corona, with weak and noisy type-C QPO detected in 901-903. Finally, a possible undetected second jet related to the second radio flare takes away the corona once again, leading to the disappearance of the QPO in 904. To verify this scenario in our data, we fit the spectra with $const*tbabs*(diskbb+bknpower)$. Unfortunately, $T_{in}$ and $\Gamma_1$ cannot be constrained in this model, due to the lack of LE part. However, we do find $\Gamma_2$ increases from $\sim$2.77 to $\sim$3.05 during the QPO disappearances in both 701 and 904, which is consistent with \cite{Sriram2021}. Additionally, the $E_{cut}$ in Table~\ref{tab:fitting} evolves from $\sim$60 keV to a very high value in both 701 and 904 illustrates that the type-C QPO is caused by a low-temperature corona \citep{Sriram2021}.

%The change of the frequency from $\sim 3$ to $\sim$10 may be related to the discrete jet. A two-component hot flow model is sometimes used to study the QPO transition \citep{Sriram2021}. The evolution in the centroid frequency may be related to the outer part ejection of the hot flow.
%During the QPO regime of 701, $E_{cut}$ is 77.27 keV. After the first radio flare start, it rises up to a very high level, indicating the sudden increasing of the corona temperature. When the radio flare ends, the corona restores its lower temperature state (QPO regime of 904), and it rises up again in the second radio flare.

%LE from disc
%The sudden increase in the QPO centroid frequency may be related to the fluctuations of the mass accretion rate. The broad band noise of the PDS may be originate from the propagation of the mass accretion rate fluctuations in the inner flow \citep{Lyubarskii1997,Arevalo2006,Ingram2011}. The increase in accretion rate will cause a over-density in the surface density. This over-density will speed up precession if it occurred in the inner flow regions, along with the rising of the X-ray flux \citep{Ingram2011,Ingram2019}. Therefore, there may be a sudden increase in the mass accretion rate right before 701, causing the QPO centroid frequency rises from $\sim$3 Hz to $\sim$10 Hz. The increased X-ray flux from 901-904 may indicate that the accretion rate is continuing increasing until the QPO disappears in 904.

%During the first radio flare, the count rate of ME and HE data also show a reflare \citep{Huang2018} between 701 and 904, but this reflare generation mechanism seems to have no correlation with the QPO generation.

Comparing the $\Gamma$ and $T_{in}$ of the QPO and non-QPO spectra of this work and the the results in \cite{Chen2022}, we notice that the QPO spectra are always harder and the disc temperatures are higher in 701 and 901-904, although the error bars of $T_{in}$ in 701 and 904 are large. We speculate that the appearance and disappearance of QPOs on the order of seconds may have a similar generation mechanism to the long-term transient QPO evolution. We find that the S-factor has a positive correlation with <CR> of ME and HE during the QPO regime for all QPO observations (see Fig.~\ref{fig:S-CR2}), suggesting that the QPO generation could be related to the hard X-ray intensity contributed by corona.

\section{Conclusion}
As a summary, based on wavelet results, we study the transient QPOs occurring during 701 and 904 in MAXI J1535$-$571. We use wavelet and PDS to study the light curves and spectral evolution of QPO and non-QPO regimes. We suggest that the transient QPOs detected in 701 and 904 are still the type-C QPOs, and the source remains in HIMS until 904. The fitting results of the energy spectra show that the spectra become softer with higher $E_{cut}$ in the non-QPO regime compared to the QPO regime. The different spectral behaviors between ObsID 144-601 and 901-903 may be related to the radio flares reported in \cite{Russell2019}, and the transient QPO disappearance may be caused by a partial ejection of the corona. The wavelet based S-factor shows a positive correlation with the mean count rate in ME and HE, but not in LE, showing that the generation of QPO is strongly correlated with photon intensity of corona. We also find that the long-term disappearance of the transient QPO in our work may have similar physical origin to the short-term appearance/disappearance of QPO as reported by \cite{Chen2022}. Applying wavelet analysis on other sources will provide further information on the QPO evolution and physical origin. More joint X-ray and radio observations during a BH LMXB outburst can give a more complete picture of the corona-radio jet/flare relation.

\section*{Acknowledgements}
We are very grateful to the referee for the useful suggestions to improve the manuscript. This work is supported by the National Key Research and Development Program of China (Grants No. 2021YFA0718500, 2021YFA0718503), the NSFC (12133007, U1838103, 11622326, U1838201, U1838202, 11903024, U1931203), and the Fundamental Research Funds for the Central Universities (No. 2042021kf0224). This work made use of data from the \textit{Insight}-HXMT mission, a project funded by China National Space Administration (CNSA) and the Chinese Academy of Sciences (CAS).

%%%%%%%%%%%%%%%%%%%%%%%%%%%%%%%%%%%%%%%%%%%%%%%%%%
\section*{Data Availability}
Data that were used in this paper are from Institute of High Energy Physics Chinese Academy of Sciences(IHEP-CAS) and are publicly available for download from the \textit{Insight}-HXMT website.
To process and fit the spectrum and obtain folded light curves, this research has made use of XRONOS and FTOOLS provided by NASA.



\bibliographystyle{mnras}
\bibliography{example} % if your bibtex file is called example.bib

%\begin{thebibliography}{99}
%\end{thebibliography}

%%%%%%%%%%%%%%%%%%%%%%%%%%%%%%%%%%%%%%%%%%%%%%%%%%

%%%%%%%%%%%%%%%%% APPENDICES %%%%%%%%%%%%%%%%%%%%%

%\appendix



%%%%%%%%%%%%%%%%%%%%%%%%%%%%%%%%%%%%%%%%%%%%%%%%%%

% Don't change these lines
\bsp	% typesetting comment
\label{lastpage}
\end{document}